\documentstyle[12pt]{article}

\begin{document}

\title{The impossible regularization of the Nambu Jona-Lasinio model with vector
interactions.}
\author{Ian J.R. Aitchison \\
Department of Physics, Theoretical Physics,\\
University of Oxford, 1 Keble Road,\\
Oxford 0X13NP, Great-Britain. \and Georges Ripka \\
Service de Physique Th\'eorique,\\
Centre d'Etudes de Saclay,\\
F-91191\ Gif-sur-Yvette Cedex, France}
\maketitle

\begin{abstract}
We show that the procedure of regularizing the real part of the euclidean
action, while leaving the imaginary part unregularized, leads to a
non-analytic and highly singular functional of the fields. It is customary
to work with an imaginary time component of the vector field, in order to
avoid regularization of the anomalous processes. We show that this procedure
is flawed by the fact that a stationary point of the action occurs for a
real,not imaginary,time component of the vector field.\ Furthermore the
action in the vicinity of the stationary point is singular. The regularized
action is thus not suitable for an evaluation of the partition function
using a saddle point method.\ We discuss proposed solutions to this problem,
as well as other regularizations.\ They all lead to practical problems.
\end{abstract}

\footnote{%
University of Oxford preprint OUTP-95-41P}

\renewcommand{\theequation}{\arabic{section}.\arabic{equation}}

\section{Introduction.}

\setcounter{equation}{0}

There have recently been several attempts to formulate a theory of solitons
composed of quarks interacting with both scalar and vector fields \cite
{Bochum,Tubingen}.\ A problem has been raised in connection with the
treatment of the time component of the vector field in the Euclidean action.
It is convenient to work with a pure imaginary time component $iV_4$ of the
vector field, with $V_4$ Hermitian.\ This yields a classical action for the
soliton, which has both real and imaginary parts. The usual regularization
procedure consists in regularizing the real part.\ The imaginary part, which
is finite, is left unregularized. The reason for doing this is
phenomenology.\ The imaginary part gives rise to anomalous processes which
are known to fit experiment when left unregularized. Suitable counterterms
may need to be added to account for anomalous processes such as $\gamma
\rightarrow 3\pi $ \cite{Arriola95}.

We show that the resulting action is a non-analytic and highly singular
functional of the fields. We show that the regularization described above is
flawed by the fact that the stationary point of the action occurs for real
time components of the vector fields (as in the Hartree approximation) and
that the action is in fact singular in the vicinity of the stationary
point.\ As a result, the regularized action is not suitable for a saddle
point evaluation of the partition function. We discuss some of the proposed
solutions of this problem.\ We show that all regularizations are fraught
with problems.

\section{The partition function expressed in terms of quark fields.}

\setcounter{equation}{0}

We consider a Nambu Jona-Lasinio lagrangian density involving scalar and
vector interactions:
\begin{equation}
L=\overline{q}\left( i\partial _\mu \gamma ^\mu -m\right) q+\frac{g_s^2}%
2\left[ \left( \overline{q}q\right) ^2+\left( \overline{q}i\gamma _5\tau
_aq\right) ^2\right] -\frac{g_v^2}2\left( \overline{q}\gamma _\mu q\right)
\left( \overline{q}\gamma ^\mu q\right)  \label{lquark}
\end{equation}
where $m$ is a mass matrix and where the coupling constants $g_s^2$ and $%
g_v^2$ have the dimension $E^{-2}$ of an inverse energy squared. More
general 4-fermion interactions can be included without altering the argument
presented in this paper.

To avoid any possible confusion with later notation, we give explicitly the
Minkowski 4-vectors in (\ref{lquark}):
\begin{equation}
x^\mu =\left( t,\stackrel{\rightarrow }{r}\right) \quad \quad \quad x_\mu
=\left( t,-\stackrel{\rightarrow }{r}\right) \quad \quad \gamma ^\mu =\left(
\beta ,\stackrel{\rightarrow }{\gamma }\right) \quad \quad \quad \gamma _\mu
=\left( \beta ,-\stackrel{\rightarrow }{\gamma }\right)
\end{equation}
Using the notation $K_a=\left( \beta ,i\beta \gamma _5\tau _a\right) $, the
action of the system, described by the lagrangian (\ref{lquark}), can be
written as follows:
\begin{equation}
I=\int_Md_4x\left( \overline{q}\left( i\partial _\mu \gamma ^\mu -m\right)
q\right) -\int_Md_4x\left( -\frac{g_s^2}2\left( q^{\dagger }K_aq\right) ^2-%
\frac{g_v^2}2\left( q^{\dagger }\stackrel{\rightarrow }{\alpha }q\right) ^2+%
\frac{g_v^2}2\left( q^{\dagger }q\right) ^2\right)  \label{actmink}
\end{equation}
where $\int_Md_4x\equiv \int dt\int d_3r$. This action (\ref{actmink}) is
often referred to as the {\em Minkowski action.}

By canonical quantization of quark fields in (\ref{actmink}) we obtain the
hermitian hamiltonian of the system:
\begin{equation}
H\left( q,q^{\dagger }\right) =\int d_3rq^{\dagger }\left( \frac{\vec \alpha
.\vec \nabla }i+\beta m\right) q+\int d_3r\left( -\frac{g_s^2}2\left(
q^{\dagger }K_aq\right) ^2-\frac{g_v^2}2\left( q^{\dagger }\stackrel{%
\rightarrow }{\alpha }q\right) ^2+\frac{g_v^2}2\left( q^{\dagger }q\right)
^2\right)  \label{hquark}
\end{equation}

The partition function of the system can be expressed in terms of the path
integral \cite{Feynman,Ripka}:
\begin{equation}
Tre^{-\beta H}\equiv \int D\left( q\right) D\left( q^{\dagger }\right)
e^{-I\left( q^{\dagger },q\right) }\quad \quad \quad I\left( q,q^{\dagger
}\right) =\int_0^\beta d\tau \left( q^{\dagger }\partial _\tau q+H\left(
q^{\dagger },q\right) \right)  \label{qpf}
\end{equation}
where the integration variables $q^{\dagger }$ and $q$ are Grassman
variables, where $H\left( q,q^{\dagger }\right) $ is the form (\ref{hquark})
and where $\beta $ is the inverse temperature. The action $I\left(
q,q^{\dagger }\right) $ is the {\em Euclidean action}. Explicitly, the path
integral (\ref{qpf}) is:
\[
Tre^{-\beta H}\equiv \int D\left( q\right) D\left( q^{\dagger }\right)
e^{-\int d_4xq^{\dagger }\left( \partial _\tau +\frac{\vec \alpha .\vec
\nabla }i+\beta m\right) q}
\]
\begin{equation}
e^{-\int d_4x\left( -\frac{g_s^2}2\left( q^{\dagger }K_aq\right) ^2-\frac{%
g_v^2}2\left( q^{\dagger }\stackrel{\rightarrow }{\alpha }q\right) ^2+\frac{%
g_v^2}2\left( q^{\dagger }q\right) ^2\right) }  \label{qpf2}
\end{equation}
and $\int d_4x\equiv \int_0^\beta d\tau \int d_3r$.

\section{Bosonized form of the Euclidean action.}

\setcounter{equation}{0}

>From the identity $\int D\left( S\right) e^{-\frac 1{2g^2}\left(
S-g^2\left( \bar q\Gamma q\right) \right) ^2}=\int D\left( S\right)
e^{-\frac 1{2g^2}S^2} $ we deduce the identity:
\begin{equation}
e^{\frac{g^2\left( \bar q\Gamma q\right) ^2}2}=\frac{\int D\left( S\right)
e^{-\frac{S^2}{2g^2}+S\left( \bar q\Gamma q\right) }}{\int D\left( S\right)
e^{-\frac{S^2}{2g^2}}}  \label{straka1}
\end{equation}

The integration variable $S$ may be chosen to be real. This identity can be
used for the first two (attractive) quartic interactions of the hamiltonian (%
\ref{hquark}). For the third (repulsive) quartic interaction we can use the
identity:
\begin{equation}
e^{-\frac{g^2\left( \bar q\Gamma q\right) ^2}2}=\frac{\int D\left( S\right)
e^{-\frac{S^2}{2g^2}+iS\left( \bar q\Gamma q\right) }}{\int D\left( S\right)
e^{-\frac{S^2}{2g^2}}}  \label{strakaphimag}
\end{equation}
which follows trivially from the identity $\int D\left( S\right) e^{-\frac
1{2g^2}\left( S-ig^2\left( \bar q\Gamma q\right) \right) ^2}=\int D\left(
S\right) e^{-\frac 1{2g^2}S^2}$. Proceeding this way, we associate the
fields $S_a,\vec V$ and $V_4$ respectively to the quark bilinear forms $%
q^{\dagger }K_aq,$ $q^{\dagger }\vec \alpha q$ and $q^{\dagger }q$ and we
write the partition function (\ref{qpf}) in the form:

\begin{equation}
Tre^{-\beta H}=\frac 1N\int D\left( q\right) D\left( q^{\dagger }\right)
D\left( S\right) D\left( \vec V\right) D\left( V_4\right) e^{-I\left(
q,q^{\dagger },S,\vec V,V_4\right) }
\end{equation}
with
\[
I\left( q,q^{\dagger },S,\vec V,V_4\right) =\int d_4xq^{\dagger }\left(
\partial _\tau +\frac{\vec \alpha .\vec \nabla }i+\beta m+K_aS_a+\vec \alpha
.\vec V+iV_4\right) q
\]
\begin{equation}
+\int d_4x\left( \frac 1{2g_s^2}S_a^2+\frac 1{2g_v^2}\left( \vec
V+V_4^2\right) \right)
\end{equation}
and:
\begin{equation}
N=\int D\left( S\right) D\left( \vec V\right) D\left( V_4\right) e^{-\int
d_4x\left( \frac 1{2g_s^2}S_a^2+\frac 1{2g_v^2}\left( \vec V+V_4^2\right)
\right) }
\end{equation}
Integration of the quark fields, which now appear in quadratic form, yields
a partition function in the form:
\begin{equation}
Tre^{-\beta H}=\frac 1N\int D\left( S\right) D\left( \vec V\right) D\left(
V_4\right) e^{-I\left( S,\vec V,V_4\right) }
\end{equation}
where the Euclidean action has the familiar form:
\begin{equation}
I\left( S,\vec V,V_4\right) =-Tr\ln D+\int d_4x\left( \frac
1{2g_s^2}S_a^2+\frac 1{2g_v^2}\left( \vec V+V_4^2\right) \right)
\label{euclact2}
\end{equation}

In (\ref{euclact2}) the Dirac operator $D$ and the Dirac hamiltonian $h$
are: 
\begin{equation}
D=\partial _\tau +h\quad \quad \quad h=\frac{\vec \alpha .\vec \nabla }%
i+\beta m+K_aS_a+\vec \alpha .\vec V+iV_4\neq h^{\dagger }  \label{eucdirac3}
\end{equation}
Because of the appearance of the term $iV_4$, the Dirac hamiltonian in (\ref
{eucdirac3}) is not hermitian. We note that we have {\em not} derived (\ref
{euclact2}) by a Wick rotation $V_0\rightarrow -iV_4$ of the time component
of a vector field.\ Nor is any ``inverse'' Wick rotation called for.

The Euclidean action (\ref{euclact2}) is usually expressed in covariant form
by redefining the following Euclidean 4-vectors: 
\[
x^\mu =x_\mu =\left( \vec r,\tau \right) \quad \quad \quad \gamma ^\mu
=\gamma _\mu =\left( \vec \gamma ,i\beta \right) \quad \quad \quad V^\mu
=V_\mu =\left( \vec V,V_4\right) 
\]
\begin{equation}
\gamma _\mu \gamma _\nu +\gamma _\nu \gamma _\mu =-2\delta _{\mu \nu }
\end{equation}
With these definitions, we have $\beta D=-i\partial _\mu \gamma _\mu
+m+\Gamma _aS_a+V_\mu \gamma _\mu $ and the effective Euclidean action (\ref
{euclact2}) acquires the familiar covariant form: 
\begin{equation}
I\left( S_a,V_\mu \right) =-Tr\ln D+\int d_4x\left( \frac
1{2g_s^2}S_a^2+\frac 1{2g_v^2}V_\mu ^2\right)  \label{euclact3}
\end{equation}
where the Dirac operator is defined to be: 
\begin{equation}
D=-i\partial _\mu \gamma _\mu +m+\Gamma _aS_a+V_\mu \gamma _\mu
\end{equation}
The equivalent actions (\ref{euclact2}) and (\ref{euclact3}) are the forms
used in Refs.\cite{Bochum,Tubingen}.

\section{The regularized Euclidean effective action.}

\setcounter{equation}{0}

We consider the proper time regularization of the fermion determinant
because most calculations of solitons performed so far have used this
regularization. The proper time regularization consists in separating the
fermion determinant $-Tr\ln D$ into real and imaginary parts $-\frac 12Tr\ln
D^{\dagger }D$ and $-\frac 12Tr\ln D/D^{\dagger }$, and in regularizing only
the real part.\ Thus the proper time regularized action is:

\[
I_\Lambda \left( S,\vec V,V_4\right) =\frac{Tr}2\int_{\frac 1{\Lambda
^2}}^\infty \frac{ds}se^{-sD^{\dagger }D}-\frac 12Tr\ln D/D^{\dagger } 
\]
\begin{equation}
+\int d_4x\left( \frac 1{2g_s^2}S_a^2+\frac 1{2g_v^2}\left( \vec
V+V_4^2\right) \right)  \label{regul2}
\end{equation}
where the Dirac operator $D$ is defined by (\ref{eucdirac3}). Whereas the
Dirac operator is a simple linear analytic function of the fields, the
operator $D^{\dagger }D$ is not.\ It is therefore not surprising to find
that different analytic continuations of the real part of the effective
action have been considered in the literature \cite{Bochum,Tubingen}. The
problems we shall discuss are caused by the separation of the action into a
regularized real part and an unregularized imaginary part. They are not
resolved by regularizing the real part by another method.

Suppose that electroweak gauge fields are added to the Dirac operator and
that we make the time component of the gauge fields pure imaginary. In the
expansion of the fermion determinant $-Tr\ln D$, the odd powers of the
imaginary time component $iV_4$ of the gauge fields arise in conjunction
with the antisymmetric tensor $\epsilon _{\mu \nu \alpha \beta }$. Such
terms give rise to the usual anomalous processes \cite{anomalous}, which are
known to fit experiment when they are calculated without regularization.
Indeed in the case of $\pi \rightarrow 2\gamma $, for example, it is known
that cut-offs, which can be as low as 800\ MeV, can induce up to 40\%
discrepancies between calculated and observed values \cite{Blin}. Thus the
use of an imaginary time component of the electroweak gauge field in the
action (\ref{regul2}) is a convenient way to ensure that the amplitudes for
anomalous processes are not cut off.

\section{A stationary point of the action.}

\setcounter{equation}{0}

We will show that a stationary point of the actions exists, in which the
fields $S$ and $\vec V$ are real, where $V_4$ is pure imaginary and where,
in addition, the fields are time independent: 
\begin{equation}
S_a\left( \vec r\right) ,\vec V\left( \vec r\right) :real\quad \quad \quad
V_4\left( \vec r\right) =iV_0\left( \vec r\right) \quad \quad V_0\left( \vec
r\right) :real  \label{point}
\end{equation}
This is in fact the form which the mean fields would have in the Hartree
approximation applied to the hamiltonian (\ref{hquark}). In the vacuum the
stationary point occurs for vanishing vector fields.\ However in solitons,
the vector field at the stationary point does not vanish. It has a shape
which is similar to the vector density $\left\langle q^{\dagger }\left( \vec
r\right) q\left( \vec r\right) \right\rangle $.

At the point (\ref{point}), the Dirac operator (\ref{eucdirac3}) is: 
\begin{equation}
D=\partial _\tau +h\quad \quad \quad h=\frac{\vec \alpha .\vec \nabla }%
i+\beta m+K_aS_a+\vec \alpha .\vec V-V_0=h^{\dagger }  \label{eucdirac1}
\end{equation}
The Dirac hamiltonian $h$ in (\ref{eucdirac1}) is hermitian.\ It has real
eigenvalues: 
\begin{equation}
h\left| \lambda \right\rangle =e_\lambda \left| \lambda \right\rangle
\end{equation}
Because the fields $S,\vec V$ and $V_0$ are time independent, the Dirac
operators $D$ and $D^{\dagger }$ can be diagonalized simultaneously in the
basis: 
\begin{equation}
D\left| \lambda ,\omega \right\rangle =\left( i\omega +e_\lambda \right)
\left| \lambda \omega \right\rangle \quad \quad \quad D^{\dagger }\left|
\lambda ,\omega \right\rangle =\left( -i\omega +e_\lambda \right) \left|
\lambda \omega \right\rangle  \label{lamow}
\end{equation}
It is easy to check that the imaginary part of the action vanishes in this
case.\ Using the basis (\ref{lamow}), the action (\ref{regul2}), at the time
independent point (\ref{point}), is finite and it can be written in the
form: 
\begin{equation}
I_\Lambda \left( S,\vec V,V_4=iV_0\right) =\frac 12\sum_{\lambda \omega
}\int_{\frac 1{\Lambda ^2}}^\infty \frac{ds}se^{-s\left( \omega ^2+e_\lambda
^2\right) }+\int d_4x\left( \frac 1{2g_s^2}S_a^2+\frac 1{2g_v^2}\left( \vec
V-V_0^2\right) \right)
\end{equation}

We now consider the expansion of the action (\ref{regul2}) in powers of the
fields, about the point (\ref{point}). The field variations are: 
\begin{equation}
S_a\rightarrow S_a+\delta S_a\quad \quad \quad \vec V\rightarrow \vec
V+\delta \vec V\quad \quad \quad V_4\rightarrow V_4+\delta V_4  \label{var1}
\end{equation}
where, for now, the increments $\delta S_a$, $\delta \vec V$ and $\delta V_4$
are all real. The corresponding variation of the Dirac operator (\ref
{eucdirac3}) is: 
\begin{equation}
D\rightarrow D+\delta S+\delta V\quad \quad \quad D^{\dagger }\rightarrow
D^{\dagger }+\delta S^{\dagger }+\delta V^{\dagger }  \label{var2}
\end{equation}
where: 
\begin{equation}
\delta S\equiv K_a\delta S_a\quad \quad \quad \delta V\equiv \vec \alpha
.\delta \vec V+i\delta V_4  \label{var3}
\end{equation}

If we use the basis (\ref{lamow}) to evaluate the traces, a straightforward
calculation will show that the first order variation of the action (\ref
{regul2}) is: 
\[
\delta I_\Lambda \left( S,\vec V,V_4\right) \equiv I_\Lambda \left( S+\delta
S,\vec V+\delta \vec V,iV_0+\delta V_4\right) -I_\Lambda \left( S,\vec
V,iV_0\right) 
\]
\[
=-\sum_{\lambda \omega }\int_{\frac 1{\Lambda ^2}}^\infty dse^{-s\left(
\omega ^2+e_\lambda ^2\right) }e_\lambda \left\langle \lambda \omega \left|
\beta \delta S+i\beta \gamma _5\tau _a\delta P_a+\vec \alpha .\delta \vec
V\right| \lambda \omega \right\rangle 
\]
\begin{equation}
+\sum_{\lambda \omega }\frac 1{i\omega +e_\lambda }\left\langle \lambda
\omega \left| i\delta V_4\right| \lambda \omega \right\rangle
\label{varact1}
\end{equation}
The first and second terms are respectively the contributions of the real
and imaginary parts of the action.\ The regularized real part contributes to
the first order variations of the fields $S$ and $\vec V,$ whereas the
imaginary part contributes to the first order variation of $V_4$. The first
order variation can be further reduced to the form: 
\[
\delta I_\Lambda \left( S,\vec V,V_4\right) =\int d_4x\delta S_a\left(
x\right) \left( \bar \rho _a\left( \vec r\right) +\frac 1{g_s^2}S_a\left(
x\right) \right) +\int d_4x\delta \vec V.\left( \vec \rho \left( \vec
r\right) +\frac 1{g_v^2}\vec V\left( x\right) \right) 
\]
\begin{equation}
\int d_4x\delta V_4\left( x\right) \left( i\rho \left( \vec r\right) +\frac
1{g_v^2}V_4\left( x\right) \right)  \label{stpts}
\end{equation}
In the expression (\ref{stpts}), $\bar \rho _a\left( \vec r\right) $ and $%
\vec \rho \left( \vec r\right) $ are the regularized densities: 
\[
\rho _a\left( \vec r\right) =\frac 12\sum_\lambda \left\langle \lambda \mid
\vec r\right\rangle K_a\left\langle \vec r\mid \lambda \right\rangle \frac{%
e_\lambda }{\left| e_\lambda \right| }n\left( \frac{e_\lambda }\Lambda
\right) 
\]
\begin{equation}
\vec \rho \left( \vec r\right) =\frac 12\sum_\lambda \left\langle \lambda
\mid \vec r\right\rangle \vec \alpha \left\langle \vec r\mid \lambda
\right\rangle \frac{e_\lambda }{\left| e_\lambda \right| }n\left( \frac{%
e_\lambda }\Lambda \right)
\end{equation}
and $\rho \left( \vec r\right) $ is the unregularized density: 
\begin{equation}
\rho \left( \vec r\right) =\sum_{e_\lambda <0}\left\langle \lambda \mid \vec
r\right\rangle \left\langle \vec r\mid \lambda \right\rangle  \label{vecden}
\end{equation}
The densities have been expressed in terms of the occupation probability $%
n\left( \frac{e_\lambda }\Lambda \right) $ which is particular to the proper
time regularization: 
\begin{equation}
n\left( \frac{e_\lambda }\Lambda \right) =\frac 1{\sqrt{\pi }}\int_{\frac{%
e_\lambda ^2}{\Lambda ^2}}^\infty \frac 1{\sqrt{x}}e^{-x}dx
\end{equation}
After subtracting the vacuum density, the vector density (\ref{vecden})
turns out to be finite in the case of a soliton. The equations (\ref{stpts})
show a posteriori that a time independent stationary point of the form (\ref
{point}) does indeed exist.

\section{The non-analytic and singular nature of the regularized action.}

\setcounter{equation}{0}

Consider a pure imaginary variation of the field $S$: 
\begin{equation}
S\rightarrow S+i\delta S\quad \quad \quad D\rightarrow D+iK_a\delta S_a\quad
\quad \quad i\delta S:imaginary
\end{equation}
and let us calculate the corresponding variation of the action (\ref{regul2}%
): 
\begin{equation}
\delta I_\Lambda =I_\Lambda \left( S+i\delta S,\vec V,V_4=iV_0\right)
-I_\Lambda \left( S,\vec V,V_4=iV_0\right)   \label{varis}
\end{equation}
We then find that the variation (\ref{varis}) of the action comes from the
unregularized imaginary part and that it is infinite. Thus to a real
variation of the field $S$ corresponds a finite variation of the action,
whereas to an imaginary variation of the field $S$ corresponds an infinite
variation of the action. Likewise, if we make an imaginary variation of the
field $V_4$, the corresponding variation of the action would come from the
regularized real part and the result would not be equal to the last term of (%
\ref{stpts}) multiplied by $i$. These examples show that the regularized
action (\ref{regul2}) is a non-analytic functional of the fields $S$ and $V_4
$ and that furthermore, it is a singular functional of the field $S$. In the
next section, we show that it is also a singular functional of the fields $%
\vec V$ and $V_4$. This is why the discussions of its analytic properties
are hazardous to say the least.

\section{A problem connected to the second order variation of the action.}

\setcounter{equation}{0}

Consider the second order variation of the action about the point (\ref
{point}). The regularized real part of the action makes a finite
contribution. The un-regularized imaginary part, however, makes an infinite
second order contribution. To see how this comes about, and to discuss a
possible solution to this problem, let us write out explicitly the second
order contribution of the unregularized imaginary part: 
\[
I_{imag}^{\left( 2\right) }\left( S,\vec V,V_4\right) = 
\]
\begin{equation}
-\frac 14Tr\frac 1{D^{\dagger }}\left( \delta S^{\dagger }+\delta V^{\dagger
}\right) \frac 1{D^{\dagger }}\left( \delta S^{\dagger }+\delta V^{\dagger
}\right) +\frac 14Tr\frac 1D\left( \delta S+\delta V\right) \frac 1D\left(
\delta S+\delta V\right)
\end{equation}
The terms which are second order in $\delta S=\delta S^{\dagger }$ cancel
out.\ This is not so for the terms which are second order in $\delta V\neq
\delta V^{\dagger }$. A straightforward calculation will show that a
non-vanishing term proportional to $\delta V_i\delta V_4$ remains which is
unregularized and infinite.

Thus, although the regularized action (\ref{regul2}) is finite at the
stationary point (\ref{point}), variations of the action around the
stationary point are infinite. This is yet another illustration of the
singular nature of the regularized action (\ref{regul2}). The problem
remains when imaginary variations of the vector fields are performed. As a
result, the regularized action (\ref{regul2}) is not suitable for a saddle
point evaluation of the partition function. It is this problem which led the
authors of Refs.\cite{Bochum,Tubingen} to modify the action (\ref{regul2}).\
It is {\em not} some fundamental neccessity of transforming the Euclidean
action back to a Minkowski action.

\section{A proposed solution to the problem.}

\setcounter{equation}{0}

Arriola and coworkers have proposed a modification to the regularized action
which can avoid this problem, while maintaining the anomalous processes
unregularized.\ The proposed solution consists in replacing, in (\ref{regul2}%
), $D^{\dagger }$ by a suitably modified Dirac operator $\bar D$. The
regularized action is then defined to be:

\[
I_\Lambda \left( S,\vec V,V_4\right) =\frac{Tr}2\int_{\frac 1{\Lambda
^2}}^\infty \frac{ds}se^{-s\bar DD}+\frac 12Tr\ln \bar D/D 
\]
\begin{equation}
+\int d_4x\left( \frac 1{2g_s^2}S_a^2+\frac 1{2g_v^2}\left( \vec
V+V_4^2\right) \right)  \label{regul3}
\end{equation}
where the Dirac operators $D$ and $\bar D$ are: 
\[
D=\partial _\tau +h\quad \quad \quad h=\frac{\vec \alpha .\vec \nabla }%
i+\beta m+K_aS_a+\vec \alpha .\vec V+iV_4 
\]
\begin{equation}
\bar D=-\partial _\tau +\bar h\quad \quad \quad \bar h=\frac{\vec \alpha
.\vec \nabla }i+\beta m+K_aS_a+\vec \alpha .\vec V-iV_4  \label{dbardef}
\end{equation}
Of course, $D$ is the same as before. As long as the field $V_4$ remains
real, we have $\bar D=D^{\dagger }$ and the regularized actions (\ref{regul2}%
) and (\ref{regul3}) are the same. However, when $V_4$ is pure imaginary,
that is, when $V_4=iV_0$, then: 
\[
D=\partial _\tau +h\quad \quad \quad h=\frac{\vec \alpha .\vec \nabla }%
i+\beta m+K_aS_a+\vec \alpha .\vec V-V_0 
\]
\begin{equation}
\bar D=-\partial _\tau +\bar h\quad \quad \quad \bar h=\frac{\vec \alpha
.\vec \nabla }i+\beta m+K_aS_a+\vec \alpha .\vec V+V_0
\end{equation}
and $\bar D\neq D^{\dagger }$. This choice has several advantages and also
disadvantages.\ 

The operator $\bar DD$ is an analytic function of the fields, whereas $%
D^{\dagger }D$ was not. The first term of the action (\ref{regul3}) is even
in $V_4$ or $V_0$, whereas the second term is odd.\ This ensures that
anomalous processes remain unregularized in the vacuum, where a stationary
point exists with vanishing vector fields. The action (\ref{regul3}) has
also disadvantages.\ When the time component $V_4$ of the vector field is
pure imaginary, the Dirac hamiltonians become hermitian, however they do not
commute.\ As a result the Dirac operators $D$ and $\bar D$ can no longer be
simultaneously diagonalized when the fields are time independent. Soliton
calculations become more complicated, because they require the
diagonalization of the three non commuting hamiltonians $h,\bar h$ and $%
h\bar h$. But, foremost, the argument $s\bar DD$ in the exponential of the
expression (\ref{regul3}) is not neccessarily positive definite.\ It would
be for small enough vector fields.\ Provisional estimates \cite{Bochum}
indicate that the operator $\bar DD$ acquires negative eigenvalues for model
parameters which are quite close to the ones required to form a hedgehog
soliton..

In Ref\cite{Tubingen} it was proposed to solve this problem by using an
effective action which includes only the second order expansion of the
action (\ref{regul2}) in powers of the vector field. This is the same as the
solution proposed in Ref.\cite{Bochum}, except that it is carried out only
to second order in the vector field. The validity of such an expansion will
obviously break down in the vicinity of the region where the operator $\bar
DD$ develops a negative eigenvalue. This would not be felt by a second order
expansion in powers of the field.

\section{Discussion of other regularizations.}

\setcounter{equation}{0}

The problems we have encountered above are essentially due to the separation
of the action into real and imaginary parts which are not regularized in the
same way.\ Some regularization procedures do not require such special action
to be taken ``by hand'' to deal with electro-weak anomalies. For example, in
some effective theories, the fermion determinant is regularized by the
appearance of non-local fields \cite{Diakonov,Ball,Cahill}. In such cases,
when the electro-weak current is coupled to a conserved vector current in a
gauge invariant way, the anomalous processes turn out to be naturally
independent of the particular form of the cut-off function, provided that
the corrections to the currents, calculated by a Noether construction, are
included \cite{cutoffind}.\ They are therefore correctly calculated without
recourse to an extra phenomenological ansatz. This should also be true of
regularizations involving 4-momentum cut-offs \cite{Weise90,Ripka92},
although this point has not been thoroughly investigated.. However there are
other problems with such regularizations.\ The calculation of solitons with
non-local fields is more complicated and longer. Furthermore, analytic
continuation of the calculated meson propagators from euclidean space-like
momenta to on-shell time-like momenta, where the theory can really be
confronted with experiment, are flawed with ambiguities, so that non-local
fields can only reliably be used in the euclidean region.

{\vskip 0.5cm}

The authors would like to thank Enrique Ruiz Arriola for numerous
enlightening discussions. One of the authors (G.R.) wishes to thank the
Oxford-Paris Programme for support without which this work would not have
been completed.

{

\end{document}